\documentclass[format=sigconf,anonymous=false,review=false]{acmart}

\AtBeginDocument{%
  \providecommand\BibTeX{{%
    \normalfont B\kern-0.5em{\scshape i\kern-0.25em b}\kern-0.8em\TeX}}}

\copyrightyear{2022} 
\acmYear{2022} 
\setcopyright{acmcopyright}\acmConference[WWW '22]{Proceedings of the ACM Web Conference 2022}{April 25--29, 2022}{Virtual Event, Lyon, France}
\acmBooktitle{Proceedings of the ACM Web Conference 2022 (WWW '22), April 25--29, 2022, Virtual Event, Lyon, France}
\acmPrice{15.00}
\acmDOI{10.1145/3485447.3512081}
\acmISBN{978-1-4503-9096-5/22/04}

\usepackage{booktabs}
\usepackage{graphicx}
\usepackage[caption=false]{subfig}
\usepackage{color}
\usepackage{algorithm}
\usepackage{algorithmic}

\usepackage{microtype}
\usepackage{bm,amsmath,amsfonts,mathtools,mathrsfs}
\usepackage{booktabs, multirow, multicol}
\usepackage{makecell}
\usepackage{stfloats}
\usepackage{arydshln}
\usepackage{pifont}

\begin{document}
\fancyhead{}

\title{UKD: Debiasing Conversion Rate Estimation via Uncertainty-regularized Knowledge Distillation}

\author{Zixuan Xu$^{ \ddagger}$, Penghui Wei$^{ \ddagger}$, Weimin Zhang,  Shaoguo Liu, Liang Wang and Bo Zheng$^{\scriptscriptstyle *}$}
\thanks{$^{ \ddagger}$Co-first authorship. $^{\scriptscriptstyle *}$Corresponding author.}
\affiliation{
  \institution{Alibaba Group} 
}
\email{{xuzixuan.xzx,wph242967,dutan.zwm,shaoguo.lsg,liangbo.wl,bozheng}@alibaba-inc.com}

\begin{abstract}
In online advertising, conventional post-click conversion rate (CVR) estimation models are trained using clicked samples. However, during online serving the models need to estimate for all impression ads, leading to the sample selection bias (SSB) issue. Intuitively, providing reliable supervision signals for unclicked ads is a feasible way to alleviate the SSB issue. This paper proposes an uncertainty-regularized knowledge distillation (UKD) framework to debias CVR estimation via distilling knowledge from unclicked ads. A teacher model learns click-adaptive representations and produces pseudo-conversion labels on unclicked ads as supervision signals. Then a student model is trained on both clicked and unclicked ads with knowledge distillation, performing uncertainty modeling to alleviate the inherent noise in pseudo-labels. Experiments on billion-scale datasets show that UKD outperforms previous debiasing methods. Online results verify that UKD achieves significant improvements.
\end{abstract}

\ccsdesc[500]{Information systems~Online advertising}

\keywords{CVR Estimation, Debiasing, Distillation, Uncertainty Modeling}

\maketitle

\section{Introduction} 
In online advertising systems, post-click conversion rate (CVR) estimation is to predict the probability of conversion after an ad click event,
and predicted CVR score is a key factor in many applications such as the ranking procedure and smart bidding. In many marketing scenarios, conversion is the ultimate goal of advertisers, and thus CVR estimation plays an important role.

\begin{figure}[t]
\centering
\centerline{\includegraphics[width=\columnwidth]{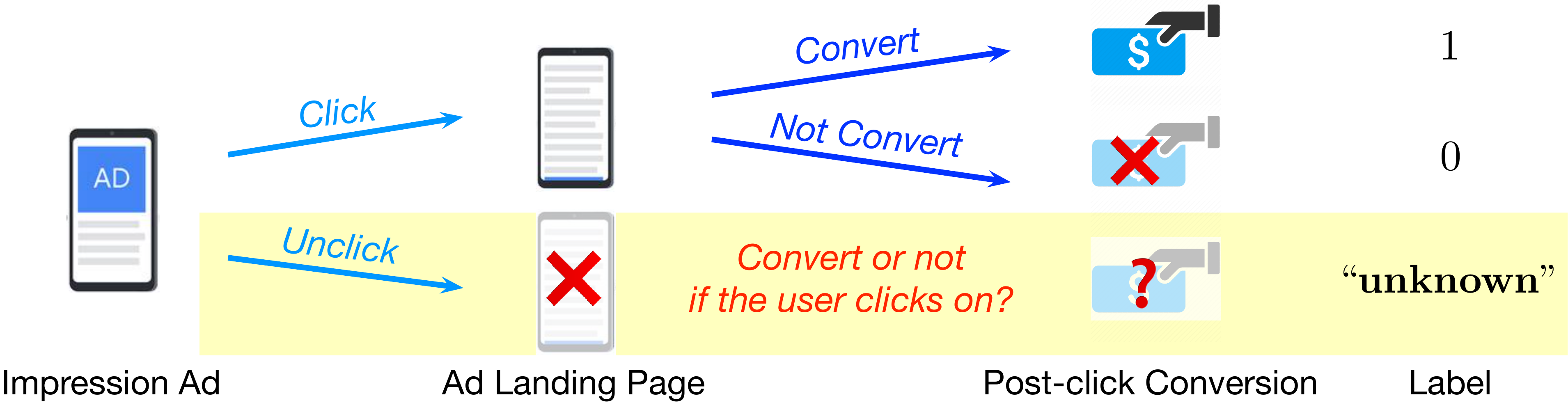}}
\vspace{-1em}
\caption{User click and conversion behaviors in advertising.}
\label{table:example}
\vspace{-1.3em}
\end{figure}

Figure~\ref{table:example} shows user click and conversion behaviors in online advertising. If users click on an impression ad, they will arrive at a landing page that shows the detailed information of the ad, and then users might take conversion actions or not. 
Obviously, only clicked ads have post-click conversion labels, and for the ads that users do \textit{not} click on, we will never know whether post-click conversion actions will happen. 
Due to the lack of ground-truth labels for unclicked ads, conventional CVR estimation models are typically trained using clicked ads only, but the models need to predict CVR for entire impression ads (including both clicked and unclicked ones) during online serving.
The problem that there is a gap between training space (i.e., \textit{click space}) and inference space (i.e., \textit{entire impression space}) is called sample selection bias (SSB)~\cite{zadrozny2004learning}.
These models may be \textit{biased} for unclicked ads, because their training procedures do not learn much knowledge from unclick ads.

The representative methods probing into the SSB issue can be divided into two categories: 1) auxiliary task learning based ~\cite{ma2018esmm,wei2021autoheri}, and 2) counterfactual learning based methods~\cite{zhang2020multidr,guo2021enhanced}. 
~\citet{ma2018esmm} propose to incorporate two auxiliary tasks that can be trained in entire impression space to indirectly learn an entire space CVR estimator. 
However, for unclicked ads, the model tends to optimize the predicted CVR scores to zeros (see §~\ref{limitations_esmm} for proof) while their actual conversion labels are ``unknown''. 
~\citet{zhang2020multidr} employ counterfactual learning to produce a theoretically unbiased CVR estimator, but the training procedure of CVR task does not explicitly take unclicked ads into account. 
In all, current methods cannot essentially debias CVR models for unclicked ads, and the SSB issue in CVR estimation is still an open problem. 

To learn entire space CVR models that can accurately estimate CVR for all impression ads, a feasible way is to provide reliable pseudo-conversion labels for unclicked ads as supervision signals. 
In this way, the training procedure of CVR models can explicitly utilize both clicked ads (with labels from logs) and unclicked ads (with pseudo-labels). Thus, these models can benefit from learning with unclicked ads compared to the ones trained on clicked ads only. 
To achieve this, the key is how to produce reliable pseudo-conversion labels for unclicked ads when we can only access ground-truth labels of clicked ads, as well as how to learn an accurate entire space CVR estimator with both ground-truth labels and pseudo-labels. 
For the former, consider that there is a discrepancy between the data distributions of clicked and unclicked ads (which causes the SSB issue)~\cite{zhang2020multidr}, we propose to formulate pseudo-conversion label generation as an unsupervised domain adaptation problem. Click/unclick space is regarded as source/target domain, and our goal is to obtain pseudo-labels for unlabeled unclicked ads (target domain) based on labeled clicked ads (source domain). 
For the latter one, consider that the confidence of unclicked ads' pseudo-labels is inferior to clicked ads' ground-truth labels, we propose to reduce the negative impact of inherent noise existed in pseudo-labels by modeling their uncertainty during  training. 

Motivated by the above considerations, in this paper we propose \textbf{U}ncertainty-regularized \textbf{K}nowledge \textbf{D}istillation (\textbf{UKD}), which aims to debias CVR estimation via distilling knowledge from unclicked ads. 
UKD's overall workflow contains a click-adaptive teacher model that produces pseudo-conversion labels for unclicked ads, and an uncertainty-regularized student model that can effectively distill the knowledge in unclicked ads learned by the teacher. 
Specifically, to produce supervision signals for unclicked ads, the teacher learns click-adaptive representations for impression ads using domain adaptation, and its predicted CVR scores on unclicked ads are taken as their pseudo-conversion labels. 
Then the student can learn from both clicked ads (with ground-truth labels) and unclicked ads (with pseudo-labels from teacher), and also performs uncertainty estimation to pseudo-labels for alleviating the inherent noise in them. 
For each unclicked ad, our student estimates its pseudo-label's uncertainty and dynamically adjust the weight of its CVR loss during training to weaken its negative impact.
Experimental results on billion-scale datasets show that \textsc{UKD} outperforms previous state-of-the-art methods. We have deployed \textsc{UKD} in Alibaba advertising platform, and online results verify that \textsc{UKD} achieves significantly improvements. 
The main contributions of this work are:

\begin{itemize}
    \item We propose uncertainty-regularized knowledge distillation (UKD) to debias CVR models via learning from unclicked ads. It employs a click-adaptive teacher to generate pseudo-conversion labels for unclicked ads, and then trains a student model that takes both clicked and unclicked ads into account. 
    \item Our student model performs uncertainty estimation to pseudo-labels generated by the teacher, alleviating the inherent noise to reduce the negative impact during distillation. 
    \item Experimental results on public and large-scale production datasets show that UKD outperforms the state-of-the-art methods. Online experiments further verify that it achieves significantly improvements on core metrics.
\end{itemize}

\section{Prerequisites}
\subsection{Problem Definition}\label{sec-preliminary} 
In online advertising systems, we can log user feedbacks on impression ads to train models for estimating CTR (click-through rate), CVR and CTCVR (click-through conversion rate).
Let $\mathcal{D}=\left\{(x, y_{click}, y_{conv},y_{{pv}\text{-}{conv}})\right\}$ denote the collected dataset of impression ads. For each impression sample, $x$ denotes the feature information, which is usually a high-dimensional vector consisting of one-hot encodings from user, ad and context fields. 
$y_{click}$, $y_{conv}$ and $y_{{pv}\text{-}{conv}}$ are the binary labels of click event, post-click conversion event and post-view conversion event respectively. 

According to the values of click labels, we divide all samples in $\mathcal{D}$ into two subsets: clicked samples $\mathcal{D}_{click}=\{\mathcal{D} \mid y_{click}=1\}$ (their conversion labels $y_{conv}$ are observed) and unclicked samples $\mathcal{D}_{unclick}=\{\mathcal{D} \mid y_{click}=0\}$ (all conversion labels are ``unknown''). 

CTR estimation is to predict the probability of click event, i.e., $p_{CTR}=p(y_{click}=1\mid x)$. 
CVR estimation is to predict the probability of conversion if a user has clicked on an ad, i.e., $p_{CVR}=p(y_{{pv}\text{-}{conv}}=1\mid y_{click}=1, x)=p(y_{conv}=1\mid x)$. And for CTCVR estimation, we have $ p_{CTCVR}=p(y_{{pv}\text{-}{conv}}=1\mid  x)=p_{CTR}\cdot p_{CVR}$.

Conventional CVR estimation models employ similar techniques as in CTR estimation task, such as logistic regression~\cite{richardson2007predicting}, factorization machines~\cite{rendle2010factorization,juan2016field} and deep neural networks (DNN) ~\cite{covington2016deep,guo2017deepfm}. Next we introduce both conventional models and entire space models. 

\subsection{Base CVR Models Trained in Click Space}

\subsubsection{\textbf{Single-Task CVR Model}}\label{model_singlecvr}
Conventional CVR estimation models are trained on the clicked data $\mathcal{D}_{click}$. 
Let $\hat{p}_{CVR}=F_v\left(x\right)$ denotes a single-task CVR model, where $\hat{p}_{CVR}\in (0,1)$ is the predicted CVR score for the impression ad $x$. $F_v(\cdot)$ represents a network that consists of a feature embedding layer and several dense layers. 
The  objective is formulated based on cross-entropy loss $\ell(\cdot, \cdot)$:
\begin{equation}
    \min_{F_v} \frac{1}{|\mathcal{D}_{click}|}\sum_{\mathcal{D}_{click}} \ell\left(y_{conv}, F_v\left(x\right)\right)\,.
\end{equation}

\subsubsection{\textbf{Joint Estimation of CVR and CTR}}\label{model_joint}
To alleviate the data sparsity issue in CVR task, jointly optimizing CVR and CTR estimation tasks is a commonly-used way, because the CTR task is trained on impression ads $\mathcal{D}$ and has much richer samples than the CVR task~\cite{ma2018esmm,wang2019deep}. The joint model contains a shared feature embedding layer, as well as two separate dense blocks to predict CVR and CTR scores respectively. Let $\hat{p}_{CVR}=F_v\left(x\right)$ and $\hat{p}_{CTR}=F_c\left(x\right)$ denote the predicted CVR and CTR scores of the joint model (note that $F_v(\cdot)$ and $F_c(\cdot)$ share the feature embedding layer), the objective of the joint model is:
\begin{equation}
    \min_{F_v,F_c}  \frac{1}{|\mathcal{D}_{click}|}\sum_{\mathcal{D}_{click}} \ell\left(y_{conv}, F_v\left(x\right)\right) +\gamma  \frac{1}{|\mathcal{D}|}\sum_{\mathcal{D}} \ell\left(y_{click}, F_c\left(x\right)\right)
\end{equation} 
where $\gamma$ is a trade-off hyperparameter.

\subsubsection{\textbf{Limitations}}
The training process of the single-task CVR model does not learn from unclicked ads, and the joint model only incorporates such information by means of the shared embeddings from CTR task. 
Thus their predicted CVR scores in unclicked space may have a non-negligible deviation because there is a discrepancy between the data distributions of click and unclick ads.

\subsection{Entire Space CVR Estimation Models}
\subsubsection{\textbf{Auxiliary Task Learning based Models}}\label{model_esmm}~\citet{ma2018esmm} incorporate two auxiliary tasks, click-through rate (CTR) and click-through conversion rate (CTCVR), that can be trained in entire impression space to indirectly learn an entire space CVR estimator. The model has the same architecture as the joint model (i.e., $\hat{p}_{CVR}=F_v\left(x\right)$ and $\hat{p}_{CTR}=F_c\left(x\right)$), while the objective is to minimize the cross-entropy loss on CTCVR and CTR estimation:
\begin{equation}
    \min_{F_v,F_c}  \frac{1}{|\mathcal{D}|}\sum_{\mathcal{D}}\biggl(  \ell\left(y_{{pv}\text{-}{conv}}, F_c(x)\cdot F_v(x)\right) +\gamma \ell\left(y_{click}, F_c(x)\right) \biggr)
\end{equation} 
where $F_c(x)\cdot F_v(x)=\hat{p}_{CTR}\cdot \hat{p}_{CVR}$ is the predicted CTCVR score. With the help of learning CTCVR estimation, the network $F_v(x)$ for CVR can learn from unclicked ads, alleviating the SSB issue. 

\subsubsection{\textbf{Counterfactual Learning based Models}}\label{model_cfl} Counterfactual learning offers a way to tackle the missing-not-at-random problem~\cite{steck2010training,dudik2011doubly,bareinboim2014recovering,schnabel2016recommendations,wang2019doubly,yuan2019improving,liu2020general,chen2021autodebias}. Several recent studies~\cite{zhang2020multidr,guo2021enhanced} employ counterfactual learning, such as inverse propensity score (IPS) and doubly robust (DR) estimators, to debias CVR estimation. An IPS-based method utilizes the predicted CTR score as propensity of the CVR loss on clicked ads to achieve an theoretically unbiased CVR estimator. The optimization objective is: 
\begin{equation}
    \min_{F_v,F_c}  \frac{1}{|\mathcal{D}_{click}|}\sum_{\mathcal{D}_{click}} \frac{1}{F_c(x)}\ell\left(y_{conv}, F_v(x)\right) +\gamma  \frac{1}{|\mathcal{D}|}\sum_{\mathcal{D}} \ell\left(y_{click}, F_c(x)\right)
\end{equation} 

A DR-based method further learns an imputation model $F_i(\cdot)$ that estimates the CVR loss of each unclicked ad. The CVR task and imputation task are  alternately trained, where the $F_i(\cdot)$ is trained on clicked data $\mathcal{D}_{click}$. Refer to~\cite{zhang2020multidr,guo2021enhanced} for details.

\subsubsection{\textbf{Limitations}}\label{limitations_esmm}
Although the auxiliary task learning based models can learn from unclicked ads with the learning of CTCVR estimation task, they have two main limitations: 
\begin{itemize}
    \item For a clicked ad (i.e., $y_{click}=1$), if its post-view conversion label $y_{{pv}\text{-}{conv}}=0$, this ad is a positive sample of CTR task as well as an negative sample of CTCVR task, which may result in gradient conflict to the two learning tasks. 
    \item For an unclicked ad (i.e., $y_{click}=0$ and $y_{conv}$ is ``unknown''), the models tend to optimize the predicted CVR scores to zeros. 
    The proof is given here:
The loss of CTCVR estimation is $\ell = - y_{{pv}\text{-}{conv}} \log (\hat{p}_{CTR} \cdot \hat{p}_{CVR}) - (1- y_{{pv}\text{-}{conv}}) \log(1 - \hat{p}_{CTR} \cdot \hat{p}_{CVR})$. For an unclicked ad whose $(y_{click}, y_{conv}, y_{{pv}\text{-}{conv}}) = (0,\text{unknown},0)$, the gradient to $\hat{p}_{CVR}$ is  $\frac{\partial \ell}{\partial \hat{p}_{CVR}} = \frac{\hat{p}_{CTR}}{1-\hat{p}_{CTR}\cdot \hat{p}_{CVR}}$. 
Note that $\hat{p}_{CTR}\in (0, 1)$ and $\hat{p}_{CVR}\in (0, 1)$, thus the gradient is always positive, which means that it tends to optimize $\hat{p}_{CVR}$ of unclicked ads to $0$, but the actual label is “unknown”. 
\end{itemize}

The counterfactual learning based models have achieved state-of-the-art performance. However, the limitations of them contain:
\begin{itemize}
    \item For IPS-based models, the training procedure of the CVR task is  on clicked data and does not explicitly take unclicked ads into account. 
    \item For DR-based models, although the imputation task is used to estimate CVR loss of each unclicked ad, its learning procedure still utilizes clicked data only, and thus the imputation task is lack of accurate supervision. 
\end{itemize}

\section{Proposed Method}
We propose an \textbf{U}ncertainty-regularized \textbf{K}nowledge \textbf{D}istillation (\textbf{UKD}) framework, which aims to debias CVR estimation via distilling knowledge from unclicked ads. 
The basic idea is that we build an entire space CVR estimation model by producing reliable pseudo-conversion labels for unclicked ads as supervision signals. 

Fig.~\ref{fig:overview} illustrates the overall workflow of \textsc{UKD}, which consists of a click-adaptive teacher model that produces pseudo-conversion labels for unclicked ads, and an uncertainty-regularized student model that distills the valuable knowledge learned by the teacher to perform entire space CVR estimation. Next, we elaborate our UKD from the details of teacher and student respectively.

\subsection{Click-Adaptive Teacher Model}
The goal of the teacher in UKD is to produce pseudo-conversion labels for unclicked ads $\mathcal{D}_{unclick}$ under the condition that we can only access ground-truth conversion labels of clicked ads $\mathcal{D}_{click}$, facilitating the entire space training of CVR estimation task. 

There is a discrepancy between the feature distributions of clicked and unclicked samples.
To possess the ability of accurate inference on unclicked ads, the teacher model may not learn feature representations specific to clicked samples, but learn click-adaptive representations. 
We propose to tackle pseudo-conversion label generation from the perspective of unsupervised domain adaptation, where the source/target domain is clicked/unclicked space, inspired by~\cite{chen2020esam}. In this way, the problem is formulated as producing reliable pseudo-conversion labels for unlabeled unclicked ads ($\mathcal{D}_{unclick}$, as target domain) given labeled clicked ads ($\mathcal{D}_{click}$, as source domain).

\subsubsection{\textbf{Click-Adaptive Representation Learning}}
Specifically, our click-adaptive teacher model adopts adversarial learning~\cite{tzeng2017adversarial,ganin2016domain} that introduces a click discriminator to mitigate inconsistent feature distributions of clicked/unclicked samples during training. 

\begin{figure*}[t]
\centering
\centerline{\includegraphics[width=1.8\columnwidth]{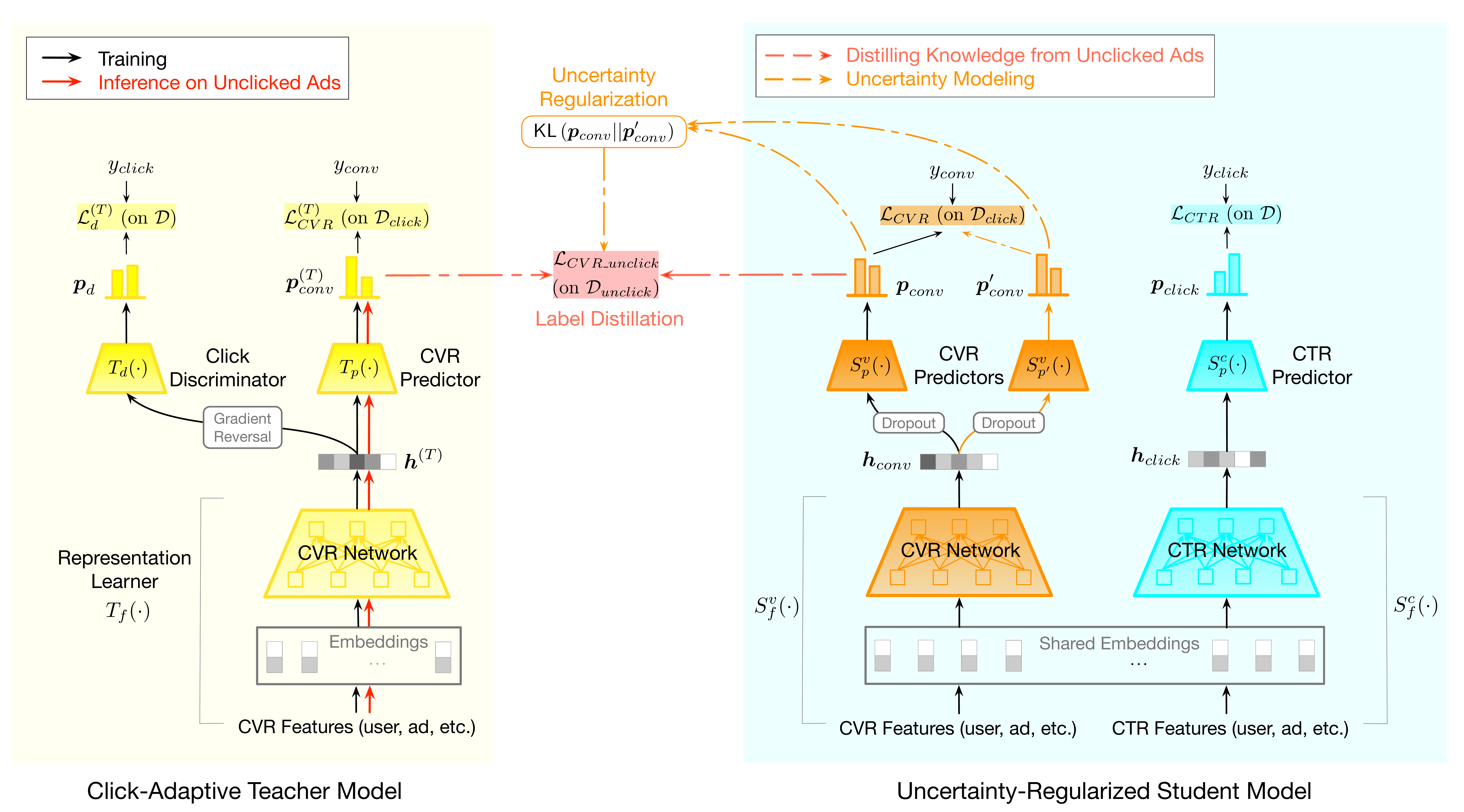}}
\vspace{-1em}
\caption{Overview of uncertainty-regularized knowledge distillation (UKD) for debiasing post-click conversion rate estimation (better viewed in color). It contains a click-adaptive teacher model that provides pseudo-conversion labels for unclicked ads, and an uncertainty-regularized student model that is trained on entire impression space.}
\label{fig:overview}
\vspace{-1em}
\end{figure*}

{\textbf{Model Architecture}}\quad As illustrated in the left part of Fig.~\ref{fig:overview}, the click-adaptive teacher model consists of a feature representation learner $T_f(\cdot)$, a CVR predictor $T_p(\cdot)$ and a click discriminator $T_d(\cdot)$. 
Formally, $T_f(\cdot)$ takes each sample's feature $x$ as input to learn its dense representation $\bm h^{(T)}$, where $T_f(\cdot)$ contains a feature embedding layer and several dense layers. $T_p(\cdot)$ intends to estimate the sample's CVR score $\hat{p}_{CVR}^{(T)}$. It consists of several dense layers and a softmax function on top to produce probability distribution.

With the aim of making the feature representation $\bm h^{(T)}$ more click-adaptive to facilitate pseudo-conversion label generation on unclicked ads, the teacher model introduces a click discriminator $T_d(\cdot)$ to classify each sample's domain (i.e., clicked or unclicked) based on the sample's representation $\bm h^{(T)}$. 
The intuition is that if a strong click discriminator cannot predict a sample's domain label correctly, its representation $\bm h^{(T)}$ is click-adaptive. 

Overall, the forward process of the teacher model is:
\begin{equation}
\begin{aligned}
    \bm h^{(T)} &= T_f(x)\\
    \bm{p}^{(T)}_{conv} &= \mathsf{softmax}\left(T_p\left(\bm h^{(T)}\right)\right)=\left(\hat{p}_{CVR}^{(T)}, 1-\hat{p}_{CVR}^{(T)}\right)\\
    \bm{p}_{d} &= \mathsf{softmax}\left(T_d\left(\bm h^{(T)}\right)\right)
\end{aligned}
\end{equation}
where $\bm{p}^{(T)}_{conv}$ is the predicted CVR distribution ($\hat{p}_{CVR}^{(T)}$ is the predicted CVR score). $\bm p_d$ is the predicted domain distribution.

\vspace{0.2em}
{\textbf{Adversarial Learning}}\quad To learn click-adaptive representations, given an impression ad $x$, its representation $\bm h^{(T)}$ learned by $T_f(\cdot)$ aims to confuse the click discriminator and maximize the domain classification loss, while the click discriminator $T_d(\cdot)$ itself aims to minimize the domain classification loss to be a strong classifier. The teacher is optimized via  an adversarial learning procedure: 
\begin{equation}
    \min_{T_f, T_p} \mathcal L^{(T)}_{CVR} = \frac{1}{|\mathcal{D}_{click}|}\sum_{\mathcal{D}_{click}} \ell(y_{conv}, \bm{p}^{(T)}_{conv})
\end{equation}
\begin{equation}
    \max_{T_f} \min_{T_d} \mathcal L^{(T)}_{d} = \frac{1}{|\mathcal{D}|}\sum_{\mathcal{D}} \ell(y_{click}, \bm{p}_{d})
\end{equation}
The first equation minimizes the loss of CVR estimation to optimize the learner $T_f(\cdot)$ and the predictor $T_p(\cdot)$. The second one means that the learner $T_f(\cdot)$ makes the representations of clicked and unclicked ads indistinguishable, while the click discriminator $T_d(\cdot)$ is optimized to better distinguish clicked ads from the unclicked ones. In practice we implement it via gradient reversal~\cite{ganin2016domain}.

The learned representations from two domains are effectively aligned when a well-trained discriminator cannot distinguish them.
Therefore, based on the click-adaptive representations, the teacher is able to make reliable CVR estimation on unclicked ads.

\subsubsection{\textbf{Produce Pseudo-Conversion Labels for Unclicked Ads}}
The trained teacher model performs inference on each unclicked ad in $\mathcal{D}_{unclick}$ to produce the predicted CVR distribution $\bm{p}^{(T)}_{conv}$ as the pseudo-conversion label, where the forward process of inference only includes $T_f(\cdot)$ and $T_p(\cdot)$, without the need of $T_d(\cdot)$. 

We use $\widetilde{\mathcal D}_{unclick}=\{(x, \bm{p}^{(T)}_{conv})\}$ to denote the unclicked samples coupled with pseudo-conversion labels, which will be utilized to train an entire space CVR model.

\subsection{Uncertainty-Regularized Student Model}
Based on unclicked ads' pseudo-conversion labels learned by the click-adaptive teacher model, our UKD framework further builds a student model based on knowledge distillation~\cite{hinton2015distilling}, which learns from both clicked ads (with ground-truth labels) and unclicked ads (with pseudo-labels) to perform entire space CVR estimation. Compared to the models that are trained using clicked samples only, our model alleviates the SSB issue via explicitly taking unclicked samples into account during training.

We elaborate distillation strategy that can guide the student model to mine the valuable knowledge learned by the teacher. 
Due to the inherent noise existed in teacher predictions, the confidence of unclicked ads' pseudo-labels is inferior to clicked ads' ground-truth conversion labels. To address this, we propose an uncertainty-regularized student that reduces the negative impact of noise by modeling pseudo-labels' uncertainty during distillation. 
Next we detail the distillation strategy of our student model with its two key modules: label distillation and uncertainty regularization.

\subsubsection{\textbf{Base Student Model: Label Distillation}}\label{model_basestudent}
We start from introducing a base student model, which is jointly learned with both CVR and CTR estimation tasks as the joint model in §~\ref{model_joint}. 

\textbf{Model Architecture}\quad The base student consists of two feature representation learners (i.e., $S_f^{v}(\cdot)$ for CVR task and $S_f^{c}(\cdot)$ for CTR task), a CVR predictor $S_p^{v}(\cdot)$ that outputs the predicted CVR score, and a CTR predictor $S_p^{c}(\cdot)$  that outputs the predicted CTR score.

Formally, the two representation learners $S_f^{v}(\cdot)$ and $S_f^{c}(\cdot)$ share the feature embedding layer, and each learner has several dense layers to learn representation $\bm{h}_{conv}$ / $\bm{h}_{click}$ w.r.t. the CVR / CTR task. Further, each of the two predictors $S_p^{v}(\cdot)$ and $S_p^{c}(\cdot)$ consists of several dense layers with a softmax function to produce probability distribution for estimating CVR / CTR score. The forward process of the base student model is:
\begin{equation}
\begin{aligned}
    \bm{h}_{conv} &= S_f^v(x),\quad \bm{h}_{click} = S_f^c(x)\\
    \bm{p}_{conv} &= \mathsf{softmax}\left(S_p^v\left({\bm h}_{conv}\right)\right) = (\hat{p}_{CVR}, 1-\hat{p}_{CVR})\\
    \bm{p}_{click} &= \mathsf{softmax}\left(S_p^c\left({\bm h}_{click}\right)\right) = (\hat{p}_{CTR}, 1-\hat{p}_{CTR})
\end{aligned}
\end{equation}

where $\bm{p}_{conv}$ denotes the predicted CVR distribution ($\hat{p}_{CVR}$ is the predicted CVR score). $\bm{p}_{click}$ and $\hat{p}_{CTR}$ can be similarly defined.

\textbf{Distilling Knowledge from Unclicked Ads}\quad With the help of unclicked ads' pseudo-conversion labels learned by the teacher, our student is optimized in entire impression space to alleviate the SSB issue. The objective of CVR estimation task is: 
\begin{equation}\label{loss_v1}
    \mathcal L_{CVR} = \underbrace{\sum_{\mathcal{D}_{click}} \ell(y_{conv}, \bm{p}_{conv})}_{\mathcal L_{CVR\_click}} + \alpha \underbrace{\sum_{\widetilde{\mathcal{D}}_{unclick}} \ell\left({\bm p}^{(T)}_{conv}, {\bm p}_{conv}\right)}_{\mathcal L_{CVR\_unclick}}
\end{equation}
where $\mathcal L_{CVR\_click}$ and $\mathcal L_{CVR\_unclick}$ are the CVR task losses on clicked and unclicked ads, and the hyperparameter $\alpha$ balances the weight of two terms.\footnote{We can distill more knowledge such as the learned representations of the representation learner $T_f(\cdot)$~\cite{romero2014fitnets,yim2017gift}. In this work we focus on verifying the effectiveness of knowledge distillation paradigm for CVR estimation, and leave that in future work.} The base student model's optimization objective is the sum of two losses about CVR task and CTR task:
\begin{equation}
    \mathcal{L}_{student} = \mathcal L_{CVR} + \gamma \mathcal L_{CTR}
\end{equation}
where $ \mathcal L_{CTR} = \sum_{\mathcal{D}} \ell(y_{click}, \bm{p}_{click}) $.

\subsubsection{\textbf{Uncertainty-regularized Student: Alleviate Noise}}\label{model_ours} It is expected that the confidence of unclicked ads' pseudo-conversion labels is inferior to that of clicked ads' ground-truth conversion labels, because the latter is obtained from user feedback logs while the former is produced by the teacher model. Due to inherent noise existed in teacher's predictions, the unclicked samples having noisy pseudo-labels mislead the student model's training procedure.

For effective knowledge distillation from unclicked ads, the key is two-fold: (i) identify noisy and unreliable unclicked samples, and (ii) reduce their negative impacts during distillation. 
To identify them, we resort to estimate the uncertainty of unclicked samples' pseudo-labels, where a higher uncertainty value indicates worse reliability. 
By using \textit{high uncertainty} as a measure of noisy unclicked samples, we can reduce the negative impacts of such samples via simply assigning \textit{low weights} to their CVR losses, which avoids misleading the student model's distillation procedure. 
Based on the above considerations, we propose an uncertainty-regularized student model. It estimates uncertainty to each unclicked ad's pseudo-label, and dynamically adjusts weights to unclicked ads' CVR losses according to uncertainty levels, reducing the negative impact of noise.~\footnote{Unlike the work~\cite{li2017learning} which aims to distill knowledge from a clean teacher, we learn from noisy pseudo-labels.}

\vspace{0.2em}
\textbf{How to Identify Noisy Samples}\quad To design a student model that possesses the ability of identifying noisy pseudo-labels, we first conduct an experiment to explore: \textit{on clean samples and noisy samples, what difference will a CVR model perform?}

We use clicked dataset $\mathcal D_{click}=\{(x, y_{conv})\}$ to run such experiment, because all labels in it are known and we can synthesize noisy dataset as well as controlling the proportion of $\frac{\#\ clean\ samples}{\#\ noisy\ samples}$. We add noise to obtain a noisy dataset $\mathcal D'_{click}$ by randomly choosing $k\%$ of positive samples in $\mathcal D_{click}$ and converting the labels from 1 to 0 (to keep the ratio of positive samples unchanged, we also convert the same size of negative samples' labels from 0 to 1). 

\begin{figure}[t]
\centering
\centerline{\includegraphics[width=0.72\columnwidth]{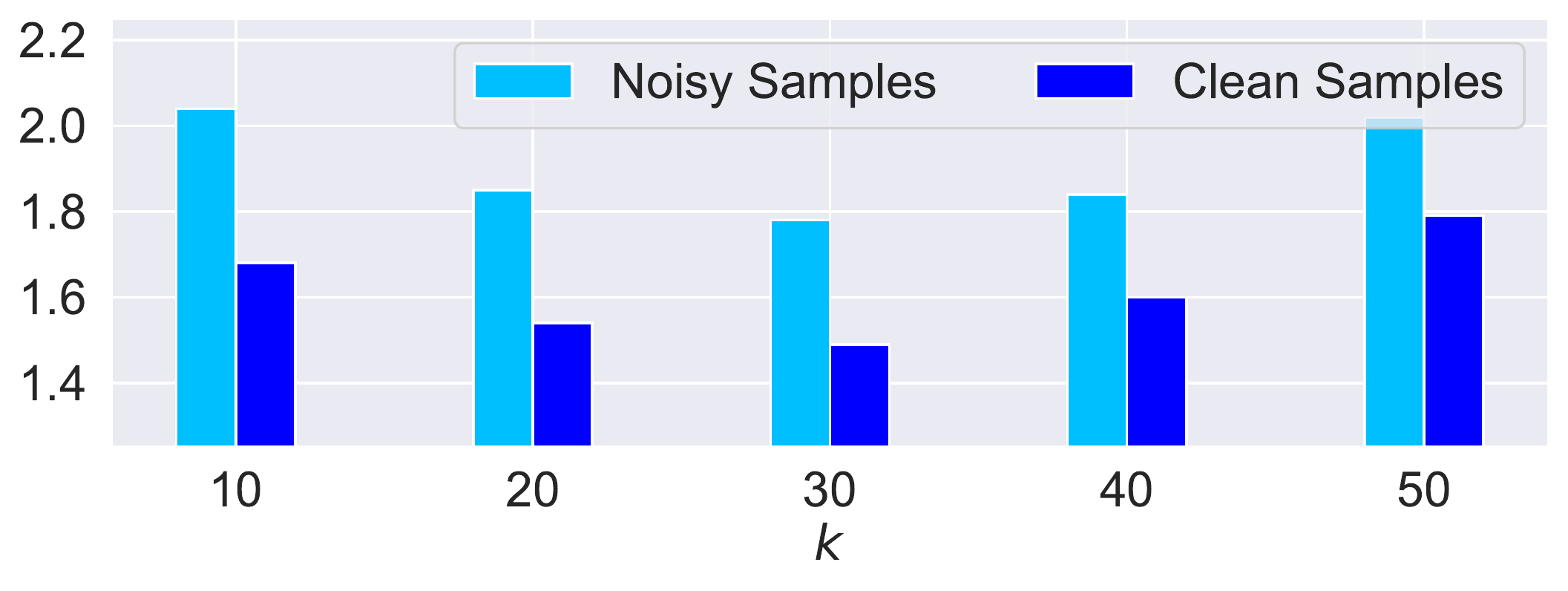}}
\vspace{-1em}
\caption{KL-divergence of two predictors' outputs ($\times 10^{-3}$).}
\label{table:2preds}
\vspace{-1em}
\end{figure}

The studies of learning from noisy data~\cite{lakshminarayanan2016simple,han2018co,shen2019regularizing,zheng2021rectifying} reveal that the inconsistency of two neural models' predictions on noisy training samples is usually larger than that on clean samples.  
Based on such guidance, we use the noisy dataset $\mathcal D'_{click}$ to train a CVR model, which contains an embedding layer, a representation learner and \textit{two seperate CVR predictors} on top (the  objective is the average of two predictors' cross-entropy losses). 
After training, we observe the averaged KL-divergence of two predictors' outputs on (1) noisy samples and (2) clean samples, respectively. Figure~\ref{table:2preds} shows the results with different $k$, and the KL-divergence on noisy samples is larger than that on clean samples for each value of $k$. 
This phenomenon can be explained that if a sample's label is clean, both two predictors easily fit the label during training, where the corresponding two loss values are small and the two predictions are similar. In contrast, if a label is noisy, the fitting procedures of two predictors will be hard to be consistent, and they tend to produce inconsistent predictions that result in a large variance. 
The experimental results verify that the inconsistency of two CVR predictors can be utilized to identify noisy training samples. 

\vspace{0.2em}
\textbf{Uncertainty Modeling}\quad Motivated by the above experiment, our uncertainty-regularized student model contains two CVR predictors $S_p^v(\cdot)$ and $S_{p'}^v(\cdot)$ to simultaneously estimate CVR scores (as illustrated in the right part of Fig.~\ref{fig:overview}), and then models the uncertainty as the inconsistency of them. Formally, let $\bm{p}_{conv}$ and $\bm{p}'_{conv}$ denote the predicted distributions from the two CVR predictors. We formulate the uncertainty as the KL-divergence of two predictions: 
\begin{equation}\label{v4_dprate}
\begin{aligned}
    & \bm{p}_{conv} = S_p^v\left(\mathsf{dropout}({\bm h}_{conv})\right)\,,\ \bm{p}'_{conv} = S_{p'}^v\left(\mathsf{dropout}'({\bm h}_{conv})\right)\\
    & \mathsf{KL}\left(\bm{p}_{conv}||\bm{p}'_{conv}\right) = \bm{p}_{conv} \log \frac{\bm{p}_{conv}}{\bm{p}'_{conv}}\\
\end{aligned}
\end{equation}
where we apply independent dropout operations to the learned representation ${\bm h}_{conv}$ to increase the discrepancy of two predictors. 

\vspace{0.2em}
\textbf{Distillation with Uncertainty Regularization}\quad Based on the estimated uncertainty for each unclicked sample, we reduce the negative impacts of noisy unclicked samples during distillation via dynamically adjusting uncertainty-based weights to CVR losses. Compared to the base student model, now the distillation procedure from unclicked ads is regularized by pseudo-label's uncertainty, alleviating the inherent noise existed in the teacher's predictions. 

For each unclicked sample, we weight its original CVR loss with a factor $\exp\left(-\lambda\cdot\mathsf{KL}\left(\bm{p}_{conv}||\bm{p}'_{conv}\right)\right)\in (0, 1]$ as uncertainty regularization. The factor is inversely related to uncertainty ($\lambda$ controls its scale). Thus, the loss $\mathcal L_{CVR\_unclick}$ is reformulated as:
\begin{equation}
        \sum_{\widetilde{\mathcal{D}}_{unclick}}\exp\left(-\lambda\cdot\mathsf{KL}\left(\bm{p}_{conv}||\bm{p}'_{conv}\right)\right)\cdot \ell\left({\bm p}^{(T)}_{conv}, {\bm p}_{conv}\right)
\end{equation}
If a sample has high uncertainty, the factor returns a small value to down-weigh its CVR loss. If the uncertainty is close to 0, the factor tends to 1, and such student devolves to the base student model. 

We also add a loss term $\sum_{\widetilde{\mathcal{D}}_{unclick}} \mathsf{KL}\left(\bm{p}_{conv}||\bm{p}'_{conv}\right) $ that acts as a regularization for uncertainty estimation. Without minimizing such term, a large KL-divergence leads to small label distillation loss, which will make the model poorly optimized.

\section{Experiments}
In this section, we conduct both offline and online experiments, and intend to answer the following research questions:
\begin{itemize}
    \item \textbf{RQ1} (offline performance): Does our proposed \textsc{UKD} outperform the state-of-the-art CVR estimation models?
    \item \textbf{RQ2} (teacher's utility): Is the teacher model necessary for debiasing CVR estimation models, and does the choice of teacher model affect the performance of our \textsc{UKD}? 
    \item \textbf{RQ3} (student's utility): Does the distillation strategy of uncertainty regularization effectively help the student model? 
    \item \textbf{RQ4} (model analysis): Does our \textsc{UKD} benefit from incorporating more unclicked samples during distillation? 
    \item \textbf{RQ5} (online performance): Does our \textsc{UKD} achieve improvements when we deploy it to industrial advertising system?
\end{itemize}

\subsection{Experimental Setup}
\subsubsection{\textbf{Datasets}} Offline experiments are conducted on a public available dataset Ali-CCP, and four large-scale production datasets from a leading advertising platform. Table~\ref{tab:dataset_statistics} lists the statistics.

\textbf{Public Dataset}\quad The Ali-CCP dataset~\cite{ma2018esmm} is the benchmark of CVR estimation, collected from user traffic logs in Taobao. 

\textbf{Production Datasets}\quad We further collect four large-scale production datasets from Alibaba advertising platform. Specifically, we collected 3-month consecutive user feedback logs from two different marketing scenarios: the first scenario is named \textit{EC}, which contains e-commerce ads for attracting potential customers. 
The second scenario is named \textit{LS}, in which the ads are about local life services. 
We organize them into four datasets: \textit{EC-Small/Large} and \textit{LS-Small/Large}. 
See Table~\ref{tab:dataset_statistics} for their statistics.

\begin{table}[t]
    \centering
    \caption{Statistics of five datasets.}
    \vspace{-1em}
    \footnotesize
    \begin{tabular}{ccccccc}
        \toprule
        Dataset &  \# Impression  & \# Click & \# Conversion & \# Time Interval \\
        \midrule
        Ali-CCP & 84M & 3.4M & 18k & N/A \\
        EC-Small  & 0.18B & 26M & 88k & 39 Days \\
        EC-Large & 0.51B & 95M & 0.3M & 72 Days\\
        LS-Small    & 0.28B & 15M & 0.2M & 31 Days\\
        LS-Large & 0.75B & 37M & 0.5M & 92 Days \\
        \bottomrule
    \end{tabular}
    \label{tab:dataset_statistics}
    \vspace{-1em}
\end{table}

\subsubsection{\textbf{Competitors of CVR Estimation}}
We compare our \textit{UKD} with the following strong baselines. Except the first model, all the rest models are trained with impression dataset $\mathcal{D}$: 
\begin{itemize}
    \item \textit{SingleCVR} (§~\ref{model_singlecvr})\quad  is a single-task network that estimates $\hat{p}_{CVR}$,  and is trained on clicked samples $\mathcal{D}_{click}$. 
    \item \textit{Joint} (§~\ref{model_joint})\quad   is a multi-task model that estimates both CVR and CTR. 
    The CTR task is trained on impression data $\mathcal{D}$, and the CVR task is trained on clicked data $\mathcal{D}_{click}$. 
    \item \textit{ESMM} (§~\ref{model_esmm})\quad   is an entire space model that learns CTR and CTCVR estimation~\cite{ma2018esmm}. 
    It is optimized in impression space $\mathcal{D}$, where the predicted CTCVR is equal to $\hat{p}_{CTR}\cdot \hat{p}_{CVR}$. 
    \item \textit{Division}\quad is a variant of \textit{ESMM}, which formulates the CVR estimation task as $p_{CVR} = p_{CTCVR} / p_{CTR}$~\cite{ma2018esmm}. 
    Compared to \textit{ESMM}, its two dense blocks produce $\hat{p}_{CTR}$ and $\hat{p}_{CTCVR}$. 
    \item \textit{CFL} (§~\ref{model_cfl})\quad   employs counterfactual learning and achieves the state-of-the-art performance~\cite{zhang2020multidr,guo2021enhanced}. Here we implement the model in~\cite{zhang2020multidr} as our competitor because we find that its performance is superior and stable among counterfactual learning based CVR models. 
\end{itemize}

\begin{table*}[t]
    \centering
    \caption{Results on four large-scale production datasets. Bold/Underlined values denote the best/second-best results.}
    \vspace{-1em}
    \footnotesize 
    \begin{tabular}{ccccccccccccc}
        \toprule 
        \multirow{2}{*}{Method} & \multicolumn{6}{c}{Dataset: EC-Small}  &  \multicolumn{6}{c}{Dataset: EC-Large}\\
        ~  & AUC$_{\text{CTCVR}}$ & AUC$_{\text{CVR}}$ & D-AUC$_{\text{CVR}}$ & NLL$_{\text{CTCVR}}$ & NLL$_{\text{CVR}}$ & D-NLL$_{\text{CVR}}$  & AUC$_{\text{CTCVR}}$ & AUC$_{\text{CVR}}$ & D-AUC$_{\text{CVR}}$ & NLL$_{\text{CTCVR}}$ & NLL$_{\text{CVR}}$ & D-NLL$_{\text{CVR}}$  \\
        \midrule
        \textit{SingleCVR} &  0.7401 & 0.6531 & 0.6558 & 0.00393 & 0.02095 & 0.02372 
        & 0.7454 & 0.6634 & 0.6623 & 0.003908 & 0.02087 & 0.02347 \\
        \textit{Joint}  &0.7445 & 0.6584 & 0.6582  & 0.00391 & 0.02091 & 0.02355 
        & 0.7470 & 0.6685 & 0.6705 & 0.003908 & 0.02086 & 0.02323 \\
        \textit{Division} & 0.7434 & 0.6559 & 0.6572 & 0.00392 & 0.02104 & 0.02371 
        & 0.7471 & 0.6635 & 0.6625 & 0.003905 & 0.02096 & 0.02335 \\
        \textit{ESMM} & 0.7441 & 0.6584 & 0.6585 & 0.00391 & \underline{0.02086} & \underline{0.02349} 
        & 0.7480 & \underline{0.6686} & 0.6697 & 0.003900 & 0.02093 & 0.02359  \\
        \textit{CFL} & \underline{0.7453} & \underline{0.6600} & \underline{0.6587} & \underline{0.00391} & 0.02110 & 0.02381 
        & \underline{0.7486} & 0.6685 & \underline{0.6722} & \underline{0.003899} & \underline{0.02067} & \underline{0.02321}  \\
        \midrule
        \textit{UKD} & \textbf{0.7513} & \textbf{0.6699} & \textbf{0.6732} & \textbf{0.00389}  & \textbf{0.02077} & \textbf{0.02347} 
        & \textbf{0.7531} & \textbf{0.6741} & \textbf{0.6752}  & \textbf{0.003890} & \textbf{0.02066} & \textbf{0.02319} \\
        \bottomrule
        \toprule
        \multirow{2}{*}{Method} & \multicolumn{6}{c}{Dataset: LS-Small}  &  \multicolumn{6}{c}{Dataset: LS-Large}\\
        ~  & AUC$_{\text{CTCVR}}$ & AUC$_{\text{CVR}}$ & D-AUC$_{\text{CVR}}$ & NLL$_{\text{CTCVR}}$ & NLL$_{\text{CVR}}$ & D-NLL$_{\text{CVR}}$  & AUC$_{\text{CTCVR}}$ & AUC$_{\text{CVR}}$ & D-AUC$_{\text{CVR}}$ & NLL$_{\text{CTCVR}}$ & NLL$_{\text{CVR}}$ & D-NLL$_{\text{CVR}}$  \\
        \midrule
        \textit{SinlgeCVR} & 0.7801 & 0.6773 & 0.6711 & 0.00413 & 0.08505 & 0.10098 
        & 0.7833 & 0.6835 & 0.6723  & 0.00412 & 0.08521 & 0.10225 \\
        \textit{Joint} & 0.7856 & 0.6927 & 0.6792 & 0.00411  & 0.08556 & 0.10174 
        & 0.7861 & 0.6911 & 0.6774  & 0.00410 & 0.08487 & 0.10241\\
        \textit{Division} & 0.7822 & 0.6851 & 0.6725  & \textbf{0.00405} & \textbf{0.08417} & \textbf{0.10001}
        & 0.7839 & 0.6741 & 0.6638 & 0.00414 & \textbf{0.08236} & \textbf{0.09993} \\
        \textit{ESMM} & \underline{0.7864} & \underline{0.6936} & \underline{0.6801} & 0.00406 & 0.08487 & \underline{0.10064}  
        & \underline{0.7876} & \underline{0.6924} & 0.6791 & \underline{0.00406} & \underline{0.08418} & 0.10205  \\
        \textit{CFL} & 0.7839 & 0.6823 & 0.6783 & 0.00409 & \underline{0.08438} & 0.10268 
        & 0.7849 & 0.6869 & \underline{0.6825} & 0.00409 & 0.08466 & 0.10271  \\
        \midrule
        \textit{UKD} & \textbf{0.7937} & \textbf{0.6958} & \textbf{0.6831}  & \underline{0.00406} & 0.08498 & 0.10086
        & \textbf{0.7955} & \textbf{0.7001} & \textbf{0.6872}  & \textbf{0.00405} & 0.08448 & \underline{0.10204}\\
        \bottomrule
    \end{tabular}
    \label{tab:offline_experiment}
    \vspace{-1em}
\end{table*}

\begin{table}[t]
    \centering
    \caption{Results on the public Ali-CCP dataset.}
    \vspace{-1em}
    \label{tab:offline_experiment_public}
    \scriptsize
    \begin{tabular}{ccccccc}
        \toprule
        Method & AUC$_{\text{CTCVR}}$ & AUC$_{\text{CVR}}$ & D-AUC$_{\text{CVR}}$ & NLL$_{\text{CTCVR}}$ & NLL$_{\text{CVR}}$ & D-NLL$_{\text{CVR}}$ \\
        \midrule
        \textit{SingleCVR}   & 0.6156 & 0.6514 & 0.6447 & 0.00211 & \textbf{0.04420} & \textbf{0.05507} \\
        \textit{Joint}       & 0.6206 & 0.6699 & 0.6644 & 0.00205 & 0.04553 & 0.05599 \\         
        \textit{Division}    & 0.6172 & 0.6647 & 0.6598 & 0.00205 & 0.04607 & 0.05667 \\
        \textit{ESMM}        & 0.6290 & 0.6711 & 0.6627 & 0.00206 & \underline{0.04493} & \underline{0.05535} \\
        \textit{CFL}   & \underline{0.6371} & \underline{0.6789} & \underline{0.6738} & \underline{0.00205} & 0.04510 & 0.05577 \\
        \midrule
        \textit{UKD}     & \textbf{0.6493} & \textbf{0.6919} & \textbf{0.6864}  & \textbf{0.00204} & 0.04553 & 0.05627 \\
        \bottomrule
    \end{tabular}
    \vspace{-1em}
\end{table}

\subsubsection{\textbf{Evaluation Metrics}} 
We use AUC and NLL (a.k.a., LogLoss) as evaluation metrics, where the former reflects ranking ability on candidates and the latter measures fitting ability of predicted scores.
Specifically, 
(i) AUC$_{\text{CVR}}$ and NLL$_{\text{CVR}}$ denote the metrics of CVR estimation on clicked samples in test set, because only clicked ads have post-click conversion labels $y_{conv}$ for evaluation.
(ii) AUC$_{\text{CTCVR}}$ and NLL$_{\text{CTCVR}}$ denote the metrics of CTCVR estimation on entire impression samples (i.e., the whole test set), because all samples have post-view conversion labels $y_{{pv}\text{-}{conv}}$ for evaluation. Following~\cite{ma2018esmm}, we use this metric to reflect the CVR model performance of alleviating the SSB issue. 
For each competitor, we compute predicted CTCVR score as $\hat p_{CTCVR}=\hat p_{CTR}\cdot \hat p_{CVR}$, where $\hat p_{CVR}$ is estimated by the competitor, and $\hat p_{CTR}$ is from the same independently trained CTR model. 

We further design two new metrics, named Debiased-AUC and Debiased-NLL (D-AUC and D-NLL for short), to evaluate the performance of entire space CVR estimation using clicked samples only. Formally, the conventional $\mathrm{AUC}_\mathrm{CVR}$ metric on clicked samples is:
\begin{equation} 
    \frac{\sum_{i\in \mathcal D^+_{click}, j\in \mathcal D^-_{click}} \mathbb{I}\left(\hat p_{CVR}(i) > \hat p_{CVR}(j)\right) }{|\mathcal D^+_{click}|\cdot |\mathcal D^-_{click}|}
\end{equation}
where $\mathcal D^+_{click}$ and $\mathcal D^-_{click}$ denote the sets of positive and negative samples respectively, $\hat p_{CVR}(i)$ is the predicted CVR score of sample $i$, and  $\mathbb{I}(\cdot)\in \{0, 1\}$ is indicator function. 
Inspired by the idea of inverse propensity score estimator~\cite{schnabel2016recommendations,saito2020unbiased}, which utilizes $\hat p_{CTR}$ as propensity score to induce an unbiased estimate of true prediction error (i.e., $\mathbb E_{\mathcal D_{click}}\left[ \frac{1}{|\mathcal D|} \sum_{\mathcal D} \frac{1}{\hat p_{CTR}} \cdot y_{click}\cdot \ell\left(y_{conv}, \hat p_{CVR}\right) \right]$), 
we assign each sample $i$ with a weight $\frac{1}{\hat p_{CTR}(i)}$ to compute D-AUC$_{\text{CVR}}$:
\begin{equation}
    \frac{\sum\limits_{i\in \mathcal D^+_{click}, j\in \mathcal D^-_{click}} \frac{1}{\hat p_{CTR}(i)}\cdot \frac{1}{\hat p_{CTR}(j)} \cdot \mathbb{I}\left(\hat p_{CVR}(i) > \hat p_{CVR}(j)\right) }{ \left(\sum_{i\in \mathcal D^+_{click}} \frac{1}{\hat p_{CTR}(i)} \right)   \cdot \left(\sum_{j\in \mathcal D^-_{click}} \frac{1}{\hat p_{CTR}(j)} \right) }
\end{equation}
We can similarly define the metric D-NLL$_{\text{CVR}}$ by modifying NLL$_{\text{CVR}}$ (equations are listed in Appendix~\ref{appendix_nll}). 

\subsection{Performance Comparison (RQ1)}
Table~\ref{tab:offline_experiment} and Table~\ref{tab:offline_experiment_public} show the comparative results of all CVR models on production and public datasets respectively. 
On all datasets, \textit{SingleCVR} performs poorly, which indicates that the models trained using clicked samples only may not be suitable for CVR estimation. 
The training procedure of the \textit{Joint} model incorporates unclicked samples via a shared embedding layer between CVR and CTR tasks, and it consistently outperforms \textit{SingleCVR}. Thus CVR estimation models can  benefit from sharing embeddings between the two tasks.

\textit{ESMM} is a strong baseline, which alleviates the SSB issue via learning CTR and CTCVR estimation in impression space as auxiliary tasks to indirectly produce an entire space CVR estimator. It outperforms \textit{Joint} on production datasets and achieves the second-best performance on the \textit{LS} datasets. This demonstrates that learning with CTCVR estimation task is an effective way to exploit unclicked samples. 
Although \textit{Division} has the same motivation with \textit{ESMM}, it performs worse than \textit{Joint}. We suggest that the reason is the instability of division operation, which may introduce  unstable numerical range that does harm to model optimization. 
\textit{CFL} is the state-of-the-art by producing a theoretically unbiased CVR estimator. 
On \textit{EC} datasets, it is superior to \textit{ESMM} on most metrics and achieves the second-best performance, indicating that counterfactual learning has advantages to alleviate the SSB issue.  

Our \textit{UKD} outperforms all competitors by a large margin on both the public and production datasets. 
Specifically, compared to the second-best results, \textit{UKD} consistently shows around 5\textperthousand\ improvement on AUC$_{\text{CTCVR}}$ and D-AUC$_{\text{CVR}}$, which is a large uplift on billion-scale dataset. This demonstrates that \textit{UKD} is an effective debiasing framework for CVR estimation, which benefits from pseudo-conversion labels learned by the click-adaptive teacher model and distillation strategy of the uncertainty-regularized student model. 

In the next three sections, we further conduct detailed experiments on \textit{EC-Small} dataset to verify the effectiveness of \textit{UKD} from three perspectives: the teacher's utility (§~\ref{rq2}), the student's utility (§~\ref{rq3}), and effects of unclicked samples (§~\ref{rq4}).

\subsection{Utility of the Click-Adaptive Teacher (RQ2)}\label{rq2}

\subsubsection{\textbf{Necessity of Click-Adaptive Teacher}}
The teacher model in \textit{UKD} is responsible to learn click-adaptive representation and produce pseudo-conversion labels for unclicked ads, aiming to alleviate the SSB issue via explicitly taking unclicked samples into account during training CVR models (i.e., the student in \textit{UKD}). 

To verify the necessity of the knowledge distillation paradigm (i.e., incorporating such a teacher model for pseudo-labels), we compare the base version of \textit{UKD} with the models that utilize unclicked samples but do not follow knowledge distillation: 
\begin{itemize}
    \item  \textit{UKD-base} (§~\ref{model_basestudent})\quad is the base version of \textit{UKD}, which contains a click-adaptive teacher model and a base student model (equation~\ref{loss_v1}) that does not utilize uncertainty modeling. 
    \item \textit{Joint+D}\quad directly embodies domain adaptation into the \textit{Joint} model by adding a click discriminator to the dense block $F_v(\cdot)$ of CVR task, without incorporating a teacher model. 
\end{itemize}

\begin{table}[t]
    \centering
    \caption{Necessity of the click-adaptive teacher. }
    \vspace{-1em}
    \small
    \begin{tabular}{cccccc}
        \toprule
        w/ teacher? & Method & AUC$_{\text{CTCVR}}$ & AUC$_{\text{CVR}}$  & D-AUC$_{\text{CVR}}$\\
        \midrule
        \checkmark & \textit{UKD-base} & \textbf{0.7485}& \textbf{0.6664}  & \textbf{0.6689} \\
        $\times$   & \textit{Joint+D} & 0.7375& 0.6464  & 0.6511 \\
        $\times$   & \textit{Joint} &0.7445 & 0.6584 &  0.6582 \\
        $\times$   & \textit{CFL}   & 0.7453& 0.6600  & 0.6587\\
        \bottomrule
    \end{tabular}
    \label{tab:KD_effectiveness}
    \vspace{-1em}
\end{table}

Results are shown in Table~\ref{tab:KD_effectiveness}. 
\textit{UKD-base} beats all other models (including the state-of-the-art model \textit{CFL}), indicating that learning click-adaptive representations for producing pseudo-conversion labels is an ideal solution for entire space CVR estimation. 
\textit{Joint+D} performs worse than our \textit{UKD-base}. According to the performance drop on AUC$_{\text{CVR}}$, the poor results can be attributed the reason that adding an discriminator hurts the optimization on clicked samples.  
The superiority of \textit{UKD-base} verifies that the knowledge distillation paradigm is necessary to alleviate the SSB issue.

\subsubsection{\textbf{Effectiveness of Click-Adaptive Teacher}}
A well-trained teacher model can provide powerful guidance for distilling unclicked samples' knowledge to the student model.

To verify the effectiveness of our click-adaptive teacher model, we compare \textit{UKD-base} to a variant that replaces the teacher from our click-adaptive model with a \textit{SingleCVR} model (i.e., a naive teacher that does not learn any information from unclicked ones) and keeps the student model unchanged (i.e., a base student model in §~\ref{model_basestudent}). 
By comparing the variant's performance with our \textit{UKD-base}, we can verify the effectiveness of the click-adaptive teacher. 
Table~\ref{tab:teacher_cvrauc} shows the comparison results. We observe that equipping our click-adaptive teacher can boost the performance on all metrics around 3\textperthousand, demonstrating the effectiveness of unsupervised domain adaptation for producing pseudo-conversion labels on unclicked ads.

We also evaluate the click discriminator in our teacher model. 
Its output $\bm p_d$ predicts the domain of an impression ad, which can be regarded as the predicted CTR distribution. We use $\bm p_d$ to calculate CTR AUC with click labels $y_{click}$. 
We observe that at both training and test phrases, CTR AUC is always around $\bm{0.50}$, indicating that the learned representations are indeed click-adaptive because they fools the well-trained click discriminator.
Thus, our click-adaptive teacher model eliminates the discrepancy between the representations of clicked and unclicked ads.

\begin{table}[t]
    \centering
    \caption{Effectiveness of the click-adaptive teacher.}
    \vspace{-1.3em}
    \small
    \begin{tabular}{cccc}
        \toprule
        Teacher & AUC$_{\text{CTCVR}}$ & AUC$_{\text{CVR}}$  &  D-AUC$_{\text{CVR}}$\\
        \midrule
        Click-adaptive Model  & \textbf{0.7485} & \textbf{0.6664} & \textbf{0.6689} \\
        SingleCVR & 0.7462 & 0.6633 & 0.6657 \\
        \hdashline[2pt/1.2pt]
        No (i.e., \textit{Joint}) & 0.7445 & 0.6584 & 0.6582 \\
        \bottomrule
    \end{tabular}
    \vspace{-0.8em}
    \label{tab:teacher_cvrauc}
\end{table}

\begin{table}[t]
    \centering
    \caption{Comparisons of different uncertainty strategies. }
    \vspace{-1.25em}
    \label{tab:uncertainty_comparison}
    \small
    \begin{tabular}{cccc}
        \toprule
        Uncertainty Strategy  & AUC$_{\text{CTCVR}}$ &  AUC$_{\text{CVR}}$ &  D-AUC$_{\text{CVR}}$\\
        \midrule
        Ours  & \textbf{0.7513}& \textbf{0.6699}  & \textbf{0.6732} \\
        Monte-Carlo dropout  & 0.7490 & 0.6678 & 0.6695  \\
        \hdashline[2pt/1.2pt] 
        No (i.e., \textit{UKD-base}) & 0.7485  & 0.6664 & 0.6689\\
        \bottomrule
    \end{tabular}
    \vspace{-0.8em}
\end{table}

\subsection{Utility of the Uncertainty-Regularized Student (RQ3)}\label{rq3}

To alleviate noisy pseudo-conversion labels produced by the teacher, our student model employs the variance of two CVR predictors to estimate uncertainty during distillation. 
To verify this strategy's effectiveness, we compare it with Monte-Carlo dropout~\cite{gal2016dropout}, a representative method for uncertainty estimation, which employs the variance of repeated predictions from the same model (but with different dropout at inference) as the uncertainty for a sample. 

To adopt Monte-Carlo dropout into our knowledge distillation framework, for each unclicked ad, the trained teacher model (after adding dropout with rate 0.2) performs inference 10 times to produce pseudo-labels. The mean of 10 predictions is used as the pseudo-label, and the variance is used as its uncertainty. We then rank all unclicked samples in ascending sort order based on their uncertainty, and retain the top 80\% samples (i.e., lower uncertainty) to train a base student model for CVR estimation. 

Table~\ref{tab:uncertainty_comparison} lists the comparison results. 
Compared to Monte-Carlo dropout, our strategy shows around 2\textasciitilde3\textperthousand\ improvement, indicating that uncertainty regularization is  more effective for alleviating label noise. 
Besides, repeated predictions in Monte-Carlo dropout consume much more computing resources (for reference, the time cost of performing CVR estimation on \textit{EC-Small} dataset is over 30 minutes, and we need to perform 10 times to estimate the uncertainty).
In contrast, our uncertainty regularization strategy only employs an additional CVR predictor $S_{p'}^v(\cdot)$ following the representation learner, thus the introduced resource consumption is negligible.

\vspace{-0.2em}
\subsection{Model Analysis of \textit{UKD} (RQ4)}\label{rq4}
We give more analysis including model performance w.r.t. the size of unclicked samples, and hyperparameter sensitivity.\footnote{Hyperparameters used in our experiments can be found in Appendix~\ref{appendix_implementation}.} 

\subsubsection{\textbf{Effect of Unclicked Samples' Size}}
In \textit{UKD}, the core intuition of alleviating the SSB issue is to explicitly incorporate unclicked samples during training. 
To show the effect of unclicked samples' size used in the teacher model, we vary the ratio of \#clicked ads : \#unclicked ads to 1:0 (no unclicked samples), 1:0.5 (less than clicked samples), 1:1 (equal to clicked samples) and 1:6 (all unclicked samples) to train different teacher models, and then train corresponding base student models to compare their performance. 

\begin{figure}[t]
\centering
\centerline{\includegraphics[width=0.91\columnwidth]{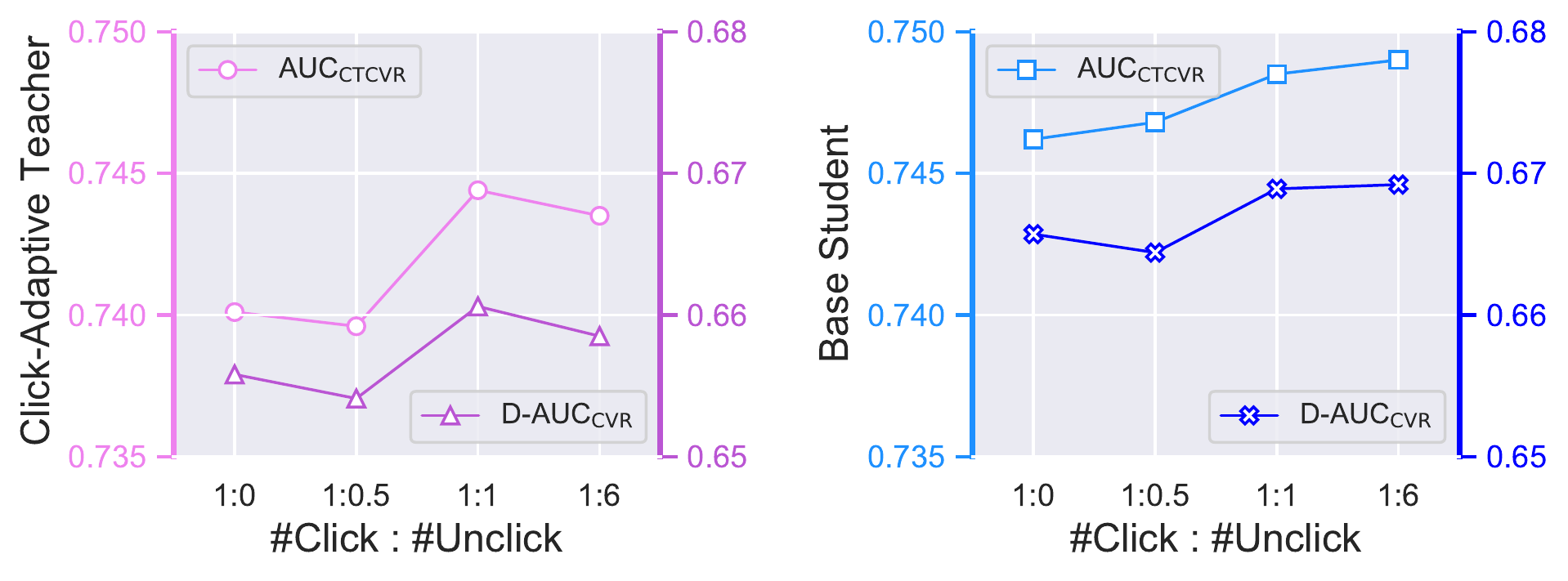}}
\vspace{-1em}
\caption{Impact of unclicked samples' size.}
\label{fig:unclicked}
\vspace{-1.5em}
\end{figure}

Figure~\ref{fig:unclicked} shows the results. As the size of unclicked samples increases, the student model's performance generally gains. Thus our \textit{UKD-base} benefits from incorporating more unclicked samples.

\subsubsection{\textbf{Hyperparameter Sensitivity Analysis}}
\begin{figure}[t]
\centering
\centerline{\includegraphics[width=0.91\columnwidth]{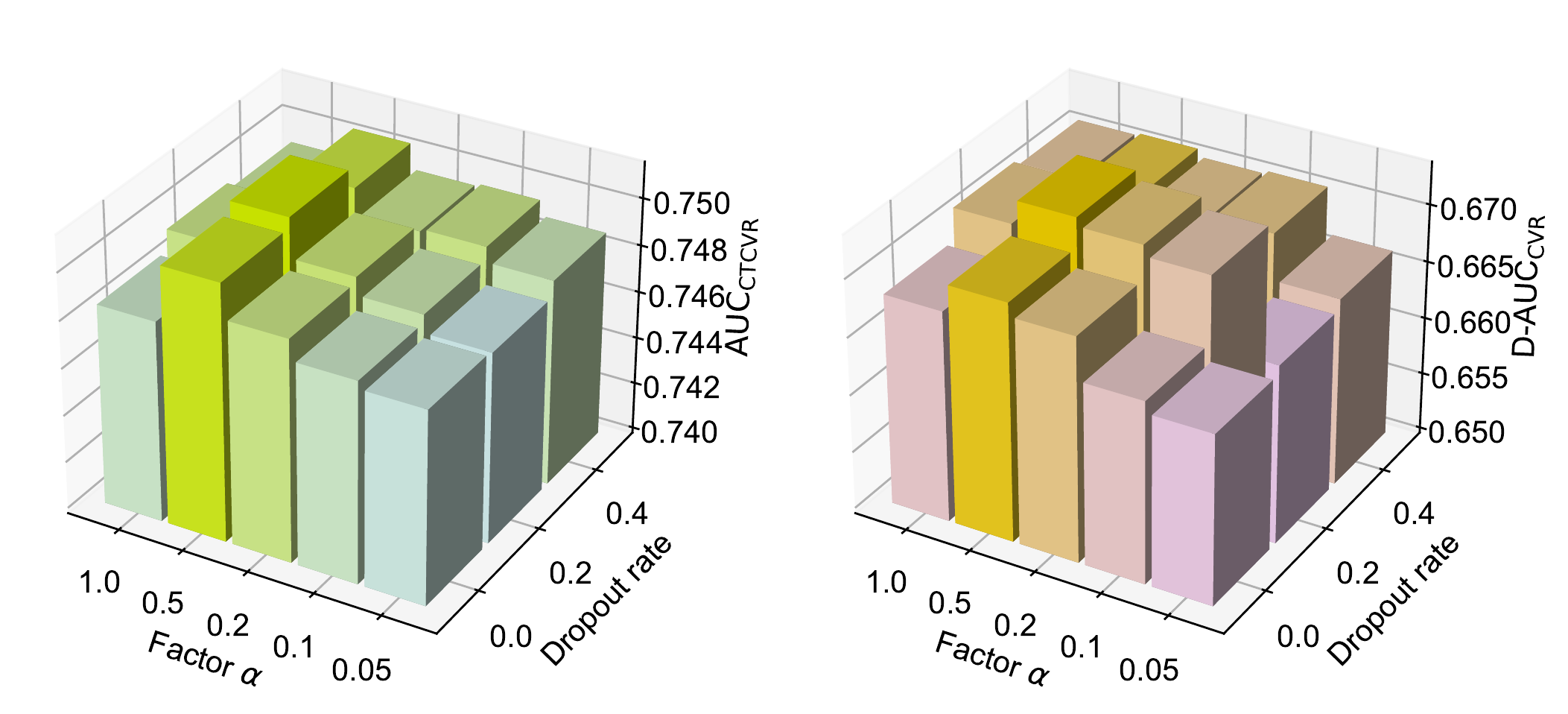}}
\vspace{-1em}
\caption{Impacts of dropout rate after the learned representation, and factor $\alpha$ on unclicked samples' losses. }
\label{fig:hyperparameter}
\vspace{-1em}
\end{figure}

In the uncertainty-regularized student model, the dropout rate after the learned representation (equation~\ref{loss_v1}) and the balance factor $\alpha$ of unclicked samples' losses (equation~\ref{v4_dprate}) are two key hyperparameters. Figure~\ref{fig:hyperparameter} illustrates the sensitivity analysis of them. 

We can observe that the model performs better when dropout rate is not 0.0, and the best result is achieved at 0.2 rate, indicating that the dropout operation is crucial to our student model. We also see that the performance usually gains with the increasing of factor $\alpha$ and the best result is achieves at $\alpha=0.5$, thus tuning the balance factor can contribute to the performance.

\vspace{-0.2em}
\subsection{Online Experimental Results (RQ5)}
We deploy \textit{UKD} to the \textit{LS} scenario on Alibaba advertising platform and conduct online A/B test for one-week.
To make a fair comparison, we follow the same configuration with the best model deployed online, such as feature set and model size. The online metrics include CVR ($\frac{\text{\#}conversion}{\text{\#}click}$), CTCVR ($\frac{\text{\#}conversion}{\text{\#}impression}$) and cost per action (i.e., CPA, equal to $\frac{total\ cost}{\text{\#}conversion}$, lower is better).

We observe that \textit{UKD} achieves \textbf{+3.4\%} lift on CVR, \textbf{+5.0\%} lift on CTCVR and \textbf{-4.3\%} lift on CPA,  
thus \textit{UKD} improves the important online metrics and promotes the performance of advertising system.

\section{Conclusion}
In this paper, we propose an uncertainty-regularized knowledge distillation framework named \textit{UKD} for debiasing CVR estimation via distilling knowledge from unclicked ads. 
It employs a click-adaptive teacher to produce pseudo-conversion labels for unclicked ads, and then trains a student model in entire space by taking both clicked and unclicked samples into account. 
Moreover, our student model performs uncertainty estimation to alleviate the inherent noise in pseudo-labels to improve the distillation performance. 
Experimental results on large-scale production datasets strongly demonstrate the superiority of \textsc{UKD} for CVR estimation. 
Online experiments further verify that it achieves significantly improvements on core online metrics including CVR, CTCVR and CPA.

\bibliographystyle{ACM-Reference-Format}
\bibliography{src-unbiasedcvr-long.bib}

\appendix
\section{Supplementary Material}
\subsection{Implementation Details}\label{appendix_implementation}
All competitors use the same feature set and embedding sizes, as well as the same model architecture for fair comparison. Each dense block contains four fully-connected layers with the output sizes of [1024, 512, 256, 2], where the first three layers belong to representation learner and the last layer belongs to predictor. During optimization, we set the batch size to 128, and adopt Adam optimizer with 0.005 learning rate.
For training the teacher model of \textit{UKD}, we randomly sample unclicked ads to keep the ratio of \#clicked ads : \#unclicked ads as 1:1 for optimizing click discriminator. For the student model, the rate of dropout operation is set to 0.2. Other hyperparameters are set as follows: $\gamma = 0.2, \alpha = 0.5, \lambda=100$. 
In practice, we treat the two predictors $S_p^v(\cdot)$ and $S_{p'}^v(\cdot)$ equally, therefore the objective $\mathcal L_{CVR}$ also contains the symmetrical terms $\sum_{\mathcal{D}_{click}} \ell(y_{conv}, {\bm p}'_{conv})$ and $\alpha \sum_{\widetilde{\mathcal{D}}_{unclick}}\exp\left(-\lambda\cdot\mathsf{KL}\left(\bm{p}'_{conv}||\bm{p}_{conv}\right)\right)\cdot \ell\left({\bm p}^{(T)}_{conv}, {\bm p}'_{conv}\right)$. 
During inference, we use the average of the two predictors' outputs for estimating CVR., i.e., $(\bm{p}_{conv}+\bm{p}'_{conv})/2$. 
For each production dataset, we use the data at the penultimate/last day as validation/test set, and all the rest data as training data.

\subsection{Definition of D-NLL$_{\text{CVR}}$}\label{appendix_nll}
Let NLL$_\mathrm{CVR}(i)$ denote the NLL of the sample $i$, and the metric NLL$_{\text{CVR}}$ is defined as:
\begin{equation}
    \frac{1}{|\mathcal D_{click}|} \sum_{i\in \mathcal D_{click}} \mathrm{NLL}_\mathrm{CVR}(i)
\end{equation}
Our metric D-NLL$_{\text{CVR}}$ is defined as: 
\begin{equation}
  \frac{1}{\sum_{i\in \mathcal D_{click}} \frac{1}{\hat p_{CTR}({i})}}\sum_{i\in \mathcal D_{click}} \frac{1}{\hat p_{CTR}({i})} \cdot \mathrm{NLL}_\mathrm{CVR}(i) \,.
\end{equation}

\end{document}